\documentclass[conference]{IEEEtran}

\usepackage{cite}
\usepackage{amsmath,amssymb,amsfonts,amsthm}
\usepackage{algorithm} 
\usepackage{algorithmicx}
\usepackage{algpseudocode}

\usepackage{graphics,
	psfrag,
	epsfig,}
 
\usepackage{tikz-cd}
\usepackage{epstopdf}

\newcommand{\cref}[1]{Constraint~\ref{#1}}

\usepackage[labelformat=simple]{subcaption}

\usepackage{graphicx}
\usepackage{textcomp}
\usepackage{xcolor}
\usepackage{tikz}
\usetikzlibrary{shapes,arrows}
\usepackage{booktabs}

\newcommand{\ignore}[1]{}
\def\BibTeX{{\rm B\kern-.05em{\sc i\kern-.025em b}\kern-.08em
    T\kern-.1667em\lower.7ex\hbox{E}\kern-.125emX}}
\begin{document}

\title{Maximizing Network  Connectivity for UAV Communications via Reconfigurable Intelligent Surfaces\\
}

\author{\IEEEauthorblockN{Mohammed S. Al-Abiad, Mohammad Javad-Kalbasi, and Shahrokh Valaee}
\IEEEauthorblockA{Department of Electrical and Computer Engineering, University of Toronto, Toronto, Canada\\ 
Email: mohammed.saif@utoronto.ca, mohammad.javadkalbasi@mail.utoronto.ca, valaee@ece.utoronto.ca
\vspace{-0.35cm}
\thanks{This work was supported in part by funding from the Innovation for Defence Excellence and Security (IDEaS) program from the Department of National Defence (DND).}
}}

\IEEEoverridecommandlockouts
\maketitle

\IEEEpubidadjcol
%\author{\IEEEauthorblockN{Mohammad Javad-Kalbasi}
%\IEEEauthorblockA{\textit{Department of Electrical Engineering} \\
%\textit{University of Toronto}\\
%Toronto, Canada \\
%mohammad.javadkalbasi@mail.utoronto.ca}
%\and
%\IEEEauthorblockN{Mohammed S. Al-Abiad}
%\IEEEauthorblockA{\textit{Department of Electrical Engineering} \\
%\textit{University of Toronto}\\
%Toronto, Canada \\
%mohammed.saif@utoronto.ca%}
%\and
%\IEEEauthorblockN{Shahrokh Valaee}
%\IEEEauthorblockA{\textit{Department of Electrical Engineering} \\
%\textit{University of Toronto}\\
%Toronto, Canada \\
%valaee@ece.utoronto.ca}}
%%%%%%%%%%%%%%%%%%%%%%%%%%%%%%%%%%%%%%%%%%%%%%%%%%%%%%%%%%%%%%%%%%%%%

%%%%%%%%%%%%%%%%%%%%%%%%%%%%%%%%%%%%%%%%%%%%%%%%%%%%%%%%%%%%%%%%%%%%%%%%%%%%%
%\maketitle

\begin{abstract}
It is anticipated that integrating unmanned aerial vehicles (UAVs) with reconfigurable intelligent surfaces (RISs), resulting in RIS-assisted UAV networks, will offer improved network connectivity  against node failures for the beyond 5G networks. In this context, we utilize a RIS to provide path diversity and alternative connectivity options for information flow from user equipment (UE) to UAVs by adding more links to the network, thereby maximizing  its connectivity. This paper employs the algebraic connectivity metric, which is adjusted by the reflected links of the RIS, to formulate the problem of maximizing the network connectivity in two cases. First, we consider formulating the problem for one UE, which is solved optimally using a linear search. Then, we consider the problem of a more general case of multiple UEs, which has high computational complexity. To tackle this problem, we formulate the problem of maximizing the network connectivity  as a semi-definite programming (SDP) optimization problem that can be solved efficiently in polynomial time. In both cases, our proposed solutions find the best combination between UE(s) and UAVs through the RIS. As a result, it tunes the  phase shifts of the RIS to direct the  signals of the UEs to the appropriate UAVs, thus maximizing the network connectivity. Simulation results are conducted to assess the performance of the proposed solutions compared to the existing solutions. 
\end{abstract}


\section{Introduction}
%\vspace{-0.20cm}
UAVs are expected to have a remarkable impact on the economy by 2026 with a global market value of US\$59.2 billion, making the incorporation of UAVs critical in beyond 5G networks \cite{UAV_economy}. One of the unique features of UAV-assisted communication is improved network connectivity by establishing line-of-sight (LoS) connections with UEs \cite{saif}. Meanwhile, RIS is a promising technique that is integrated with UAVs to further improve network connectivity \cite{M}, particularly in networks that experience deep fade.  In this context, RISs can be leveraged to provide path diversity and alternative connectivity solutions for information flow from UEs to UAVs in RIS-assisted UAV networks.

The prime concern of UAV communications is that UAV nodes  are prone to failure due to several reasons, such as limited energy, hardware failure, or targeted failure in the case of battlefield surveillance systems. Such UAV failures cause  network disintegration, and consequently, information flow from UEs to a fusion center through UAVs can be severely impacted. Hence, it is crucial to always keep the  network connected, which was addressed in the literature by adding more backhual links to the network, e.g., \cite{H}. In spite of recent advances in wireless sensor networks, most of the existing studies consider routing solutions with the focus more on extending the battery lifetime of sensor nodes. These works define network connectivity as network lifetime, in which the first node or all the nodes have failed \cite{1331424, 926982}.
However, none of the aforementioned works has ever explicitly considered the exploitation of RISs to add more reflected links for improving network connectivity. Different from works \cite{1331424, 926982} that focused on routing solutions, this paper focuses on designing a more connected RIS-assisted UAV network that enables information flow from the UEs to the UAVs even if some of the UAVs  have failed. %Thus, a more resilient network that can stay connected despite some nodes failures. 

The algebraic connectivity \cite{new}, also called  the Fiedler metric or the second smallest eigenvalue of the Laplacian matrix representing a graph, is a metric that measures how well a graph  is connected.  In the literature, such metric is usually associated with network connectivity \cite{8292633, 4657335, 4786516}. In \cite{8292633}, the authors maximized the algebraic connectivity by positioning the UAV to maximize the connectivity of small-cells backhaul network. A more general study in \cite{4657335} proposed different network maintenance algorithms to  maximize the connectivity of wireless sensor networks. Since the algebraic connectivity is a good measure of how connected the graph is, the more edges that exist between the UEs and the UAVs, the more resilient network can be designed without being disconnected due to node failures \cite{4657335, 4786516}. To this end, this paper aims to utilize the RIS to add link redundancy to the network and tune the RIS phase shift configurations to direct UEs' signals to appropriate UAVs, so that the connectivity of RIS-assisted UAV networks is maximized. To the best of our knowledge, the problem of maximizing the network connectivity in RIS-assisted UAV networks has not been studied before in the literature. 

In this paper, we address this problem by employing the concept of 
algebraic connectivity \cite{new} of a graph in network connectivity, then we consider two problem cases. First, we formulate the problem for one UE and one RIS and solve it optimally via a linear search. Then, we formulate the problem for a more general case of multiple UEs and one RIS. It is shown that solving this general problem optimally is computationally prohibitive since it requires computing the algebraic connectivity of the resulting network for  each possible edge that connects the UEs to the UAVs through the RIS. To tackle this problem, we adjust the algebraic connectivity metric of the original graph network by the candidate edges between the UEs and the UAVs via the RIS. Then, we reformulate  the problem of maximizing the network connectivity as a semi-definite programming (SDP) optimization problem that can be solved efficiently in polynomial time. In both cases, our proposed solutions find the best combination between the UE(s) and the UAVs through the RIS by tuning its  phase shifts  to direct the UEs' signals to the appropriate UAVs, thus maximizing the network connectivity. Simulation results are conducted to assess the performance of the proposed solutions compared to the existing solutions.

%The rest of this paper is organized as follows. In Section
%II, we describe the system model and the network connectivity. In Section III, we formulate the problem of maximizing the network connectivity in two different cases, and our proposed solutions are developed in Section IV. Section V presents some simulation results on the performance of the proposed solutions. Finally, we  conclude the paper in  Section VI.

\section{System Model and Network Connectivity}\label{S}
\begin{figure}[t!]
			\begin{center}
				\includegraphics[width=0.65\linewidth, draft=false]{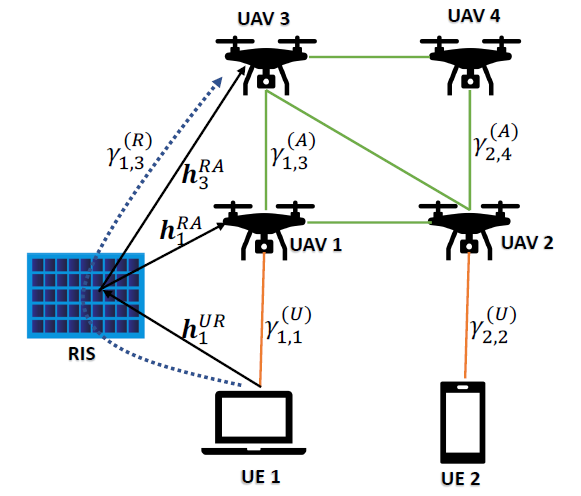}
				\caption{A typical RIS-assisted UAV network with one RIS, 2 UEs, and 4 UAVs.}
     \vspace{-5.5mm}
			\end{center}
   \label{fig2}
		\end{figure}
  
\subsection{System Model}\label{SA}
We consider a RIS-assisted UAV network with a set of UAVs, one RIS, and multiple UEs that represent ground users, sensors, etc. An example of the considered network is shown in Fig. 1. The sets of UAVs and UEs are denoted as $\mathcal A=\{1, 2, \ldots, A\}$  and $\mathcal U=\{1, 2, \ldots, U\}$, respectively, where $A$ is the cardinality of the set $\mathcal A$. All UEs and UAVs are quipped with single antennas. 
The $A$ UAVs fly and hover over
assigned locations at a fixed flying altitude and connect $U$ UEs with the fusion center. The locations of the UAVs, UEs, and the RIS are assumed to be fixed. We assume that all channels follow a quasi-static flat-fading model and thus remain constant over one time slot. The RIS is installed with a certain altitude $z_R$. Let $(x_R, y_R)$ be the 2D location of the RIS, $(x_a, y_a, z_a)$ be the 3D location of the $a$-th UAV, 
and $(x_u, y_u)$ be the 2D location of the $u$-th UE, respectively. The distances between the $u$-th UE and the RIS and between the RIS and the $a$-th UAV are denoted by $d^\text{UR}_{u}$ and $d^\text{RA}_{a}$, respectively.

Due to their altitude, UAVs can have good  connectivity to UEs. However, UEs may occasionally experience deep fade. To overcome this problem and further improve network connectivity, we propose to utilize a RIS to impose link redundancy to the RIS-assisted UAV network. As such, the network becomes more resilient against node failures by providing path diversity and alternative connectivity options between UEs and UAVs. The RIS is equipped with a controller and $M_r \times M_c$ passive reflecting units (PRUs) to form a uniform passive array (UPA). Each column of the UPA has $M_r$ PRUs with an equal spacing of $d_c$ meters (m) and each row of the UPA consists of $M_c$ PRUs with an equal spacing of $d_r$ m. These PRUs can add indirect links between UEs and UAVs with adjustable phase shifts. The phase-shift matrix of the RIS is modeled as the diagonal matrix $\mathbf{\Theta} =diag(e^{j\theta_{1}},\ldots, e^{j\theta_{2}},\ldots,e^{j\theta_{M}})$, where $\theta_{m}\in [0,2\pi)$, for $m=\{1, \ldots, M\}$ and $M = M_r \times M_c$.

The successful communications between the UEs and the RIS are measured using the distance threshold $D_o$, i.e., the $u$-th UE is connected to the RIS with distance $d^{(R)}_{u}$ if $d^{(R)}_{u} \leq D_o$.
The communications between the UEs and UAVs/RIS are assumed to occur over different time slots (i.e., time multiplexing access) to avoid interference among the scheduled UEs. Therefore, we assume that only one UE is transmitting in each time slot to reduce interference. Considering the interference among the different scheduled UEs to the RIS and the UAVs is left for future work. 

 Since this paper focuses on the network connectivity from data link-layer viewpoint, we abstract the physical layer factors and consider a model that relies only on the distance between the nodes. Therefore, we model only the large scale fading and ignore the small scale fading. To quantify the UEs transmission to the UAVs and the RIS, we use the signal-to-noise ratio (SNR). For the $u$-th UE, SNR is defined as follows \cite{8292633}
\begin{equation}
\gamma^{(U)}_{u,a}=\frac{d_{u,a}^{-\alpha}p}{N_0},
\end{equation}
where $d_{u,a}$ is the distance between the $u$-th UE and the $a$-th UAV, $p$ is the transmit power of the $u$-th UE, which is maintained fixed for all the UEs, $N_0$ is the additive white Gaussian noise (AWGN) variance, and $\alpha$ is the path loss exponent that depends on the transmission environment.

UAVs hover at high altitudes, thus we reasonably  assume that they maintain LoS channel between each other. The path loss between the $a$-th and the $a'$-th UAVs can be expressed as
\begin{equation}
\Gamma_{a,a'}=20\log\bigg( \frac{4 \pi f_c d_{a,a'}}{c} \bigg),
\end{equation}
where $d_{a,a'}$ is the distance between the $a$-th UAV and the $a'$-th UAV, $f_c$ is the carrier frequency, and $c$ is light speed. The SNR in dB between the $a$-th UAV and the $a'$-th UAV is $\gamma^{(A)}_{a,a'}=10\log P-\Gamma_{a,a'}-10 \log N_0$, where $P$ is the transmit power of the $a$-th UAV, which is maintained fixed for all the UAVs. Note that the SNR of the $u$-th UE determines whether it has a successful connection to the corresponding UAV $a$. In other words, the $a$-th UAV is assumed to be within the transmission range of the $u$-th UE if $\gamma^{(U)}_{u,a} \geq \gamma^\text{UE}_{0}$, where $\gamma^\text{UE}_{0}$ is the minimum SNR threshold for the communication links between the UEs and the UAVs.  Similarly, we assume that UAV $a$ and UAV $a'$ have a successful connection provided that $\gamma^{(A)}_{a,a'} \geq \gamma^\text{UAV}_{0}$, where $\gamma^\text{UAV}_{0}$ is the minimum SNR threshold for the communication links between the UAVs. 

Since the RIS is deployed in the higher altitude, the signal propagation of UE-to-RIS link is adopted to be a simple yet reasonably accurate LoS channel model \cite{9293155}. The LoS channel vector between the $u$-th UE and the RIS is given by \cite{9293155}
\begin{equation}
\mathbf h^\text{UR}_{u}=\sqrt{\frac{\beta_0}{(d_{u}^\text{UR})^2}}\tilde{\mathbf h}^\text{UR}_{u},
\end{equation}
where $d_{u}^\text{UR}$ is the distance between the $u$-th UE and the RIS, $\beta_0$ denotes the path loss at the reference distance $d_\text{ref}=1$ m, and $\tilde{\mathbf h}^\text{UR}_{u}$ 
represents the array response component which can be denoted by
\begin{eqnarray}\nonumber
\tilde{\bold{h}}^\text{UR}_{u}&=&{[1,e^{-j\frac{2\pi d_{r}}{\lambda}\phi_{u}^\text{UR}\psi_{u}^\text{UR} },\ldots,e^{-j\frac{2\pi d_{r}}{\lambda}(M_{r}-1)\phi_{u}^\text{UR}\psi_{u}^\text{UR} }]}^{T}\\\nonumber
&~&\otimes {[1,e^{-j\frac{2\pi d_{c}}{\lambda}\varphi_{u}^\text{UR}\psi_{u}^\text{UR} },\ldots,e^{-j\frac{2\pi d_{c}}{\lambda}(M_{c}-1)\varphi_{u}^\text{UR}\psi_{u}^\text{UR} }]}^{T},
\end{eqnarray}
where $\phi_{u}^\text{UR}, \varphi_{u}^\text{UR}$, and $\psi_{u}^\text{UR}$ are related to the sine and cosine terms of the vertical and horizontal angles-of-arrival (AoAs)
at the RIS \cite{9293155}, and given by $\phi_{u}^\text{UR}=\frac{y_{u}-y_{R}}{\sqrt{(x_u-x_R)^2+(y_u-y_R)^2}}$, $\varphi_{u}^\text{UR}=\frac{x_{R}-x_{u}}{\sqrt{(x_u-x_R)^2+(y_u-y_R)^2}}$, $\psi_{u}^\text{UR}=\frac{-z_{R}}{d^\text{UR}_{u}}$, $\lambda$ is the wavelength, and $T$ denotes transpose. On the other hand, the RIS and UAVs are deployed in the higher altitudes, thus the reflected signal propagation of the RIS-to-UAV link typically occurs in clear airspace where the obstruction or reflection effects diminish. The LoS channel vector between the RIS and the $a$-th UAV is given by
\begin{equation}
\mathbf h^\text{RA}_{a}=\sqrt{\frac{\beta_0}{(d_{a}^\text{RA})^2}}\tilde{\mathbf h}^\text{RA}_{a},
\end{equation}
where $d_{a}^\text{RA}$ is the distance between the RIS and the $a$-th UAV, and $\tilde{\mathbf h}^\text{RA}_{a}$ 
represents the array response component which can be denoted by
\begin{eqnarray}\nonumber
\tilde{\bold{h}}^\text{RA}_{a}&=&{[1,e^{-j\frac{2\pi d_{r}}{\lambda}\phi_{a}^\text{RA}\psi_{a}^\text{RA} },\ldots,e^{-j\frac{2\pi d_{r}}{\lambda}(M_{r}-1)\phi_{a}^\text{RA}\psi_{a}^\text{RA} }]}^{T}\\\nonumber
&~&\otimes {[1,e^{-j\frac{2\pi d_{c}}{\lambda}\varphi_{a}^\text{RA}\psi_{a}^\text{RA} },\ldots,e^{-j\frac{2\pi d_{c}}{\lambda}(M_{c}-1)\varphi_{r,a}^\text{RA}\psi_{a}^\text{RA} }]}^{T},
\end{eqnarray}
where $\phi_{a}^\text{RA}, \varphi_{a}^\text{RA}$, and $\psi_{a}^\text{RA}$ are related to the sine and cosine terms of the vertical and horizontal angles-of-departure (AoDs)
from the RIS to the $a$-th UAV \cite{9293155}, and respectively given by $\phi_{a}^\text{RA}=\frac{y_{R}-y_{a}}{\sqrt{(x_R-x_a)^2+(y_R-y_a)^2}}$, $\varphi_{a}^\text{RA}=\frac{x_{R}-x_{a}}{\sqrt{(x_R-x_a)^2+(y_R-y_a)^2}}$, and $\psi_{a}^\text{RA}=~\frac{z_R-z_a}{d^\text{RA}_{a}}$.

Given the aforementioned channel models, the concatenated channel for the
UE-RIS-UAV link between the $u$-th UE and the $a$-th UAV through the RIS is given by 
$h^\text{URA}_{u,a} =  (\mathbf h^\text{RA}_{a} \mathbf)^H \mathbf{\Theta} \mathbf h^\text{UR}_{u}$ \cite{9293155}. Accordingly, the SNR of the reflected link between the $u$-th UE and the $a$-th UAV through the RIS can be written as 
$\gamma^{(R)}_{u,a}= \frac{p |h^\text{URA}_{u,a}|^2}{N_0}$ \cite{9593204}.
For successful connection between UE $u$ and UAV $a$ via RIS $r$, $\gamma^{(R,r)}_{u,a} \geq \gamma^\text{RIS}_{0}$, where $\gamma^{(RIS)}_{0}$ is the minimum SNR threshold for the communication links between the UEs and the UAVs via the RISs.

We model the considered RIS-assisted UAV network as an undirected
 graph $\mathcal G(\mathcal V, \mathcal E)$, where $\mathcal V = \{v_1, v_2, \cdots, v_V\}$ is the
set of nodes (i.e., UAVs and UEs) in the network, $\mathcal E= \{e_1, e_2, \cdots, e_E\}$ is the set of all edges.
$V=|\mathcal U \cup \mathcal A| =|\mathcal V|$ and $|\mathcal E|=E$ are the numbers of vertices and edges in the graph, respectively. The graph $\mathcal G$ implies that all the links in the network are bidirectional, i.e., a node $v$ is able to reach node $v'$, and vice versa. The edge between any two nodes is created based on a typical SNR threshold.

\subsection{Network Connectivity}
%This subsection briefly discusses the definition of the Laplacian matrix $\mathbf L$ representing a graph $\mathcal G(\mathcal V, \mathcal E)$, its second smallest eigenvalue $\lambda_2(\mathbf L)$, and the relationship between $\lambda_2(\mathbf L)$ and the connectivity of the associated graph $\mathcal G(\mathcal V, \mathcal E)$.  

For an edge $e_{k}$, $ 1 \leq k \leq E$, that connects two nodes $\{v_n, v_m\} \in \mathcal V$, let $\mathbf a_k$ be a vector, where the $n$-th and $m$-th elements in $\mathbf a_k$ are given by $a_{k,n}=1$ and $a_{k,m}=-1$, respectively, and zero otherwise. The incidence matrix $\mathbf A \in \mathbf R^{V\times E}$ of a graph $\mathcal G$ is the matrix with the $k$-th column given by $\mathbf a_k$.  Hence, in undirected graph $\mathcal G(\mathcal V, \mathcal E)$, the Laplacian matrix $\mathbf L$ is an $V$ by $V$ matrix, which is defined as follows \cite{4657335}:
\begin{equation}
\mathbf L= \mathbf A \mathbf A^T=\sum^{E}_{k=1} \mathbf a_k \mathbf a^T_k,
\end{equation}
where the entries of $\mathbf L$ are given as follows:%, where $D_{v_i}$ is the degree of node $v_i$, which represents the number of the neighboring nodes of node $v_i$. $L_{i,j}=-1$ if $i,j \in \mathcal E$, otherwise $L_{i,j}=0$
\begin{equation}
L(n,m) = \begin{cases}
D_{v_n} &\text{if} ~v_n=v_m,\\
-1 &\text{if}~ (v_n, v_m) \in \mathcal E \\
0 & \text{otherwise}
\end{cases}
\end{equation}
where $n, m \in \{1, 2, \ldots, V\}$ are the indices of the nodes, and $D_{v_n}$ is the degree of node $v_n$, which represents the number of all its neighboring nodes. 

In network connectivity, algebraic connectivity, also called the Fiedler metric or the second smallest eigenvalue \cite{new}, measures how well a graph $\mathcal G$ that has the associated Laplacian matrix $\mathbf L$ is connected. From its name, this metric is usually denoted as $\lambda_2(\mathbf L)$. 
The motivation of $\lambda_2(\mathbf L)$ to be used as a network connectivity metric comes from the following two main reasons \cite{new}. First, $\lambda_2 (\mathbf L) > 0$ if and only if $\mathcal G$ is connected, i.e., $\mathcal G$ is only one connected graph. It is worth mentioning that when $\lambda_2(\mathbf L)=0$, the graph is disconnected in which at least one of its vertices is unreachable from any other vertices in the graph. Second,  $\lambda_2 (\mathbf L)$ is monotone increasing in the edge set, i.e., if $\mathcal G_1=(V, E_1)$ and $\mathcal G_2=(V, E_2)$ and $E_1 \subseteq E_2$, then $\lambda_2 (\mathbf L_2) \geq \lambda_2 (\mathbf L_1)$. This implies that $\lambda_2(\mathbf L)$ qualitatively represents the connectivity of a graph in the sense that the larger $\lambda_2(\mathbf L)$ is, the more connected the graph will be. To this end, since $\lambda_2 (\mathbf L)$ is a good measure of how connected the graph is, the more edges that exist between the UEs and the UAVs, the longer the network can live without being disconnected due to node failures. Thus, the  network becomes more  resilient. Based on that, we consider $\lambda_2 (\mathbf L)$ as a quantitative measure of the network resiliency in this paper, similar to \cite{4657335, new_lifetime}. %This direct relation between $\lambda_2 (\mathbf L)$ and the network resiliency will be validated in Section \ref{NR}. %Note that the first smallest eigenvalue $\lambda_1(\mathcal G)$ is always zero in connected graphs \cite{new, 4657335}.

%Based on the applications in wireless networks, there are many definitions of the network resiliency. For example, in \cite{1331424, 925682}, the network resiliency was related to network lifetime which is defined as the time until the first node in the network fails. In contrast, in \cite{926982}, the network resiliency was defined as the lifetime of a network until all the nodes fail. It is well-known that the network connectivity is a very important characteristic in wireless networks resiliency (e.g., ad-hoc and terrestrial networks). Therefore, it should be taken into account in the network resiliency. For example, in sensor network applications or terrestrial applications, the time until the first node fails may not serve as a good definition of the network connectivity, since the failure of the first node  may not significantly affect the information delivery/collection. In contrast, network disintegration typically causes severe impact in the information delivery. To this end, since $\lambda_2 (\mathbf L)$ is a good measure of how connected the graph is, the more edges that exist between the UEs and the UAVs, the longer the network can live without being disconnected due to node failures. Thus, the  network becomes more  resilient. Based on that, we consider $\lambda_2 (\mathbf L)$ as a quantitative measure of the network resiliency in this work, similar to \cite{4657335, new_lifetime}. %This direct relation between $\lambda_2 (\mathbf L)$ and the network resiliency will be validated in Section \ref{NR}. 

\section{Problem Formulation}

%The problem of  maximizing the connectivity  in RIS-assisted UAV networks can be stated as follows. 
Given a RIS-assisted UAV network represented by a graph $\mathcal G(\mathcal V, \mathcal E)$, what are the optimum combinations between the UEs and the UAVs through the RIS in order to maximize $\lambda_2 (\mathbf L)$ of the resulting network? 
Essentially, adding the RIS to the network may result in connecting multiple UEs to multiple UAVs, which were not connected together. It may also result in adding new alternative options to the UEs if their scheduled UAVs have failed. In this context, we leverage the RIS to add more links to the network, and by adjusting its phase shifts, RIS can smartly beamform the signals of the UEs to suitable UAVs to maximize the network connectivity. %Therefore, adding the RIS may result in adding many edges to the original graph.%, where one edge is added in each time slot.

With RIS deployment, a new graph $\mathcal G'(\mathcal V, \mathcal E')$ is constructed with the same number of $V$ nodes and a larger set of edges denoted by $\mathcal E'$ with $\mathcal E'=\mathcal E \cup e^{R}_{u,a}$, where $e^{R}_{u,a}$ is the new edge connecting the $u$-th UE to the $a$-th UAV through the RIS and $\mathcal E \subseteq \mathcal E'$. Note that the effect of deploying the RIS appears only in the edge set $\mathcal E$, and  not in the node set $V$ \cite{8292633, 4657335, 4786516}. By adding those new links to the network, the gain can be realized by computing $\lambda_2 (\mathbf L') \geq \lambda_2 (\mathbf L)$, where $\lambda_2 (\mathbf L')$ is the resulting Laplacian matrix of a graph $\mathcal G'(\mathcal V, \mathcal E')$.

We  consider that in each time slot only one UE can transmit to the RIS, which directs the UE's signal to only one UAV. In what follows, we consider two different cases of network configurations to formulate the optimization problem of maximizing $\lambda_2 (\mathbf L')$ in each time slot.\\\\
\textbf{Case 1: One UE and One RIS}

Let $\mathcal A_0$ be a set of reachable UAVs that have indirect communication links from the UE through the RIS, i.e., $\mathcal A_0 =\{a\in \mathcal A\backslash \mathcal A_{UE} \mid \gamma^{(R)}_{a} \geq \gamma^\text{RIS}_{0}\}$, where $\mathcal A_{UE}$ is the set of UAVs that have direct links to the UE. Our aim is to provide an alternative link to connect the UE to a single UAV in the set $\mathcal A_0$. As such, the UE does not miss the communication if its scheduled UAV has failed. Now, let $y_a$ be a binary variable that is equal to $1$ if the RIS is connected to the $a$-th UAV ($a \in \mathcal A_0$), and zero otherwise. The considered optimization problem in this case is formulated as follows:
\begin{subequations}
\label{eq9}
\begin{align}
&\max_{\substack{y_a, \theta_m}}  \lambda_2(\mathbf L')
\label{eq9a} \\
 &{\rm subject~to\ } 
\sum_{a \in \mathcal A_0}y_a=1, \label{eq9c}\\
& \theta_m \in [0, 2\pi)~~~~~~~~~~~~~~~~~~~~~~~~~~~ m=\{1, \ldots, M\}, \label{eq9d}\\
& y_a \in \{0, 1 \} ~~~~~~~~~~~~~~~~~~~~~~~~~~~~~~~~~~~~~~\forall a \in \mathcal A_0,
\end{align}
\end{subequations}
where constraint (\ref{eq9c}) assures that the RIS directs the signal of the UE to only one UAV. Constraint (\ref{eq9d}) is for the RIS phase shift optimization.\\ \\
\textbf{Case 2: Multiple UEs and One RIS}

Unlike case 1 that considers one UE, case 2 adds an optimization variable that selects the $u$-th UE that transmits in each time slot.  Let $\mathcal A^u_0$ be a set of reachable UAVs that have indirect communication links from the $u$-th UE through the RIS, i.e., $\mathcal A^u_0 =\{a\in \mathcal A\backslash \mathcal A_{u} \mid \gamma^{(R)}_{u,a} \geq \gamma^\text{RIS}_{0}\}$, where $\mathcal A_{u}$ is the set of UAVs that have direct links to the $u$-th UE.
Let  $x_u$ be a binary variable that is equal to $1$ if the $u$-th UE is connected to the RIS, and zero otherwise. Now, let  $y^u_a$ be a binary variable that is equal to $1$ if the RIS is connected to the $a$-th UAV when the $u$-th UE is selected to transmit, and zero otherwise. The considered optimization problem in this case is formulated as follows:
\begin{subequations}
\label{eq10}
\begin{align}
&\max_{\substack{x_u, y^u_a, \theta_m}}  \lambda_2(\mathbf L')
\label{eq10a} \\
 &{\rm subject~to\ } \sum_{u \in \mathcal U}x_u=1, \label{eq10c}\\
& \sum_{a \in \mathcal A^u_0}y^u_a=1, \label{eq10d}\\
& \theta_m \in [0, 2\pi)~~~~~~~~~~~~~~~~~~~~~~~~~~~ m=\{1, \ldots, M\}, \label{eq10f}\\
& x_u \in \{0,1 \}, y^u_a \in \{0, 1 \}  ~~~~~~~~~~~~~~~~~~~~~~~\forall a \in \mathcal A_0^u,
\end{align}
\end{subequations} 
where constraint (\ref{eq10c}) and (\ref{eq10d}) assure that only one UE can transmit to the RIS and the signal of that UE is reflected from the RIS to only one UAV.  Constraint (\ref{eq10f}) is for the RIS phase shift optimization.

%\vspace{-0.35cm}
\section{Proposed Solutions}
%\vspace{-0.25cm}
It is computably affordable to optimally solve (\ref{eq9}) since it considers only one UE, however solving (\ref{eq10}) optimally for the case of multiple UEs is computationally prohibitive. Therefore, this section proposes to solve (\ref{eq9}) optimally using a linear search over all the possible UAV nodes. Then, the section formulates (\ref{eq10}) as an SDP optimization problem that can be solved efficiently in polynomial time. The process of those proposed solutions are explained in subsections IV-A and IV-B, respectively.

\subsection{Solution of Case 1}
To optimally solve (\ref{eq9}), we consider a linear search scheme that computes  $\lambda_2(\mathbf L')$ for each possible UAV node $a \in \mathcal A_0$, and then calculate the corresponding phase shift of the RIS to that UAV node. In particular, the corresponding phase shift at PRU of the RIS to the $a$-th UAV is calculated  as follows \cite{9293155}
\begin{multline} \label{phase} 
\theta_{m}= \pi \frac{f_c}{c}\bigg\{ d_r(m_r-1)\psi^\text{RA}_{a}   \phi^\text{RA}_{a}+d_c(m_c-1)\psi^\text{RA}_{a} \varphi^\text{RA}_{a}
\\+d_r(m_r-1)\psi^\text{UR}\phi^\text{UR}  +d_c(m_c-1)\psi^\text{UR}\varphi^\text{UR}\bigg\}.
\end{multline}

We argue that the computational complexity of the proposed solution of this case is affordable since it needs to compute only $|\mathcal A_0|$ Laplacian matrices. The steps of the proposed method are summarized in Algorithm 1.

\begin{algorithm}[t!]
	\caption{The Proposed Linear Search for Case 1}
	\label{Algorithm2}
	\begin{algorithmic}[1]
		\State \textbf{Input:} One UE $u$, one RIS, $\mathcal A$ and network topology.
        \State Construct $\mathcal G(\mathcal V, \mathcal E)$.
        \For{$a=1, 2, \ldots, |\mathcal A_0|$}
        \State Calculate $\theta_m$ of the RIS to the $a$-th UAV as given in (11), $\forall m\in \{1, 2, \ldots, M\}$
        \State Set $\mathcal G' \leftarrow \mathcal G(\mathcal V, \mathcal E \cup \{e^{R}_{u,a}\})$
         \State Calculate $\lambda_2 (\mathbf L')$ of a graph $\mathcal G'$
         \EndFor         
	\State \textbf{Output:} Optimal $\lambda_2(\mathbf L')$.
	\end{algorithmic}
\end{algorithm}

\subsection{Solution of Case 2}
The exhaustive search scheme to solve (\ref{eq10}) for multiple UEs can be done by computing $\lambda_2 (\mathbf L')$ for a total of $U \sum_{u=1}^U |\mathcal A^u_0|$ Laplacian matrices, which requires huge amount of computation for large $U$. For graph $\mathcal G'(\mathcal V, \mathcal E')$,  the time complexity of the exhaustive search is high for large network settings. It runs in $\mathcal O(4 E' V^3/3)$ to compute $\lambda_2(\mathbf L')$ \cite{L}. To overcome such computational intractability, we instead propose an efficient method to solve (\ref{eq10}), which finds the feasible UE-RIS-UAV association to maximize $\lambda_2 (\mathbf L')$ using SDP solvers \cite{SDP-M}.

We add a link connecting the $u$-th UE to the $a$-th UAV through the RIS if both $x_u$ and $y^u_a$ in (\ref{eq10}) are $1$. Let $\mathbf z$ be a vector representing the UE-RIS-UAV candidate associations, in which case $x_u=1$ and $y^u_a=1$, $\forall u \in \mathcal U, a \in \mathcal A$. Therefor, the problem in (\ref{eq10}) can be seen as having a set of $|\mathbf z|$ UE-RIS-UAV candidate associations, and we want to select the optimum UE-RIS-UAV association among these
$|\mathbf z|$ associations. This optimization problem can be formulated as
\begin{align} 
&\max_{\mathbf z} \lambda_2(\mathbf L'(\mathbf z))
\label{eq10a} \\
& {\rm subject~to\ } \mathbf 1^T \mathbf z=1, \mathbf z \in \{0,1\}^{|\mathbf z|}, \nonumber
\end{align}
where $\mathbf 1 \in \mathbf R^{|\mathbf z|}$ is the all-ones vector and
\begin{equation}
\mathbf L'(\mathbf z)=\mathbf L+\sum^{|\mathbf z|}_{l=1} z_l \mathbf a_l \mathbf a^T_l,
\end{equation}
where $\mathbf a_l$ is the incidence vector resulting from adding link $l$ to the original graph $\mathcal G$ and $\mathbf L$ is the Laplacian matrix of the original graph $\mathcal G$. Clearly, the dimension of $\mathbf L$ and $\mathbf L'(\mathbf z)$ is $V \times V$.

The optimization vector in (12) is the vector $\mathbf z$. The $l$-th element of $\mathbf z$, denoted by $z_l$, is either
$1$ or $0$, which corresponds to whether this UE-RIS-UAV association should be chosen or not, respectively. The combinatorial optimization problem in (12) is NP-hard problem with high complexity. Therefore, we relax the constraint on the entries of $\mathbf z$ and allow them to take any value in the interval $[0, 1]$. Specifically, we relax the Boolean constraint $\mathbf z \in \{0,1\}^{|\mathbf z|}$ to be a linear constraint $\mathbf z \in [0,1]^{|\mathbf z|}$, then we can represent the problem (12) as
\begin{align}
&\max_{\mathbf z} \lambda_2(\mathbf L'(\mathbf z))
\label{eq10a} \\
& {\rm subject~to\ } \mathbf 1^T \mathbf z=1, 0 \leq \mathbf z \leq 1.\nonumber
\end{align}
We emphasize that the optimal solution of the relaxed problem in (14) is an upper bound for the optimal solution of the original problem (12) since it has a larger feasible set.  In \cite{4786516}, it was shown that $\lambda_2(\mathbf L'(\mathbf z))$ in (14) is the point-wise
infimum of a family of linear functions of $\mathbf z$, which is a concave function in $\mathbf z$. In addition, the relaxed constraints are linear in $\mathbf z$. Therefore, the optimization problem in (14) is a convex optimization problem \cite{4786516}, and it is equivalent to the following SDP optimization problem \cite{CON}
%\begin{subequations}
%\label{eq14}
\begin{align}
&\max_{\mathbf z, q} q
\label{eq10a} \\
& {\rm subject~to\ } q(\mathbf I - \frac{1}{|\mathbf z|}\mathbf 1 \mathbf 1^T) \preceq \mathbf L'(\mathbf z), \mathbf 1^T \mathbf z=1, 0 \leq \mathbf z \leq 1, \nonumber
\end{align}
%\end{subequations} 
where $\mathbf I \in \mathbf R^{V \times V}$ is the identity matrix and $\mathbf F \preceq \mathbf L$ denotes that $\mathbf L- \mathbf F$ is a positive semi-definite matrix.

By solving the SDP optimization problem in (15) efficiently using any SDP standard solver such as the CVX software package \cite{SDP-M}, the optimization variable $\mathbf z$ is obtained.  Since the entries of the output vector $\mathbf z$ are continuous, we consider to round the maximum entry to $1$ while others are rounded to zero. For the given association vector $\mathbf z$, we optimize the phase shift of the RIS as in (11) to direct the signal of the selected UE to the associated UAV.

%\vspace{-0.25cm}
\section{Numerical Results}\label{NR}
%\vspace{-0.25cm}
For the numerical evaluations, we use the same  RIS configurations and UAV communications that were used in \cite{8292633} and \cite{9293155}, respectively. We consider a RIS-assisted UAV system in an area of $150 ~m \times 150 ~m$, where the RIS has a fixed location and the UEs and the UAVs are distributed randomly. The considered simulation parameters are as follows: the RIS  is located at ($35~m, 50~m$) with an altitude of $20$ m,  $M=100$, $d_r=5$ cm, $d_c=5$ cm, $\beta_0=10^{-6}$, $N_0=-130$ dBm, the altitude of the UAVs is $50$ m, $f_c=3\times10^9$ Hz, $c=3\times 10^8$ m/s, $\alpha=4$, $p=1$ watt, $P=5$ watt, $\gamma_0^\text{(U)}=85$ dB, and $\gamma_0^\text{(A)}=80$ dB. Unless specified otherwise, $A=7$, $U=10$, and $\gamma^\text{(RIS)}_0=30$ dB.  

For the sake of numerical comparison, the proposed schemes are compared with the following schemes: 1) original  benchmark scheme without RIS deployment, 2) random scheme that selects a random link to connect the UE to one of the UAVs through the RIS.  For completeness of our work, we also compare the proposed SDP scheme of case $2$  with the optimal scheme that is considered as a performance upper bound since it searches over all the possible links between the UEs and the UAVs. In the simulations, the network connectivity is calculated over $500$ iterations, and the average value is presented. In each iteration, we change the locations of the UEs and the UAVs.

\begin{figure}[t!]
			\begin{center}
				\includegraphics[width=0.95\linewidth, draft=false]{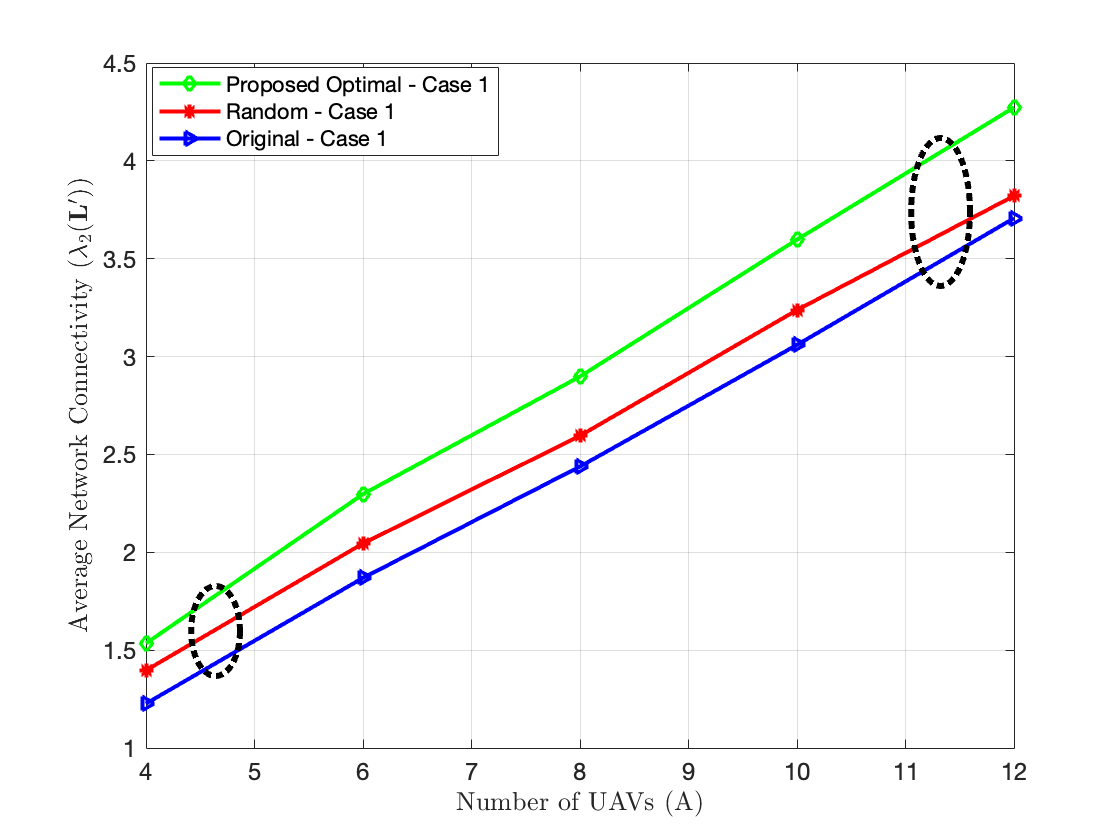}
				\caption{The average network connectivity $\lambda_2(\mathbf L')$ of case 1 versus the number of UAVs $A$.}
     \vspace{-5.5mm}
			\end{center}
   \label{fig2_case1}
		\end{figure}

\begin{figure}[t!]
			\begin{center}
				\includegraphics[width=0.95\linewidth]{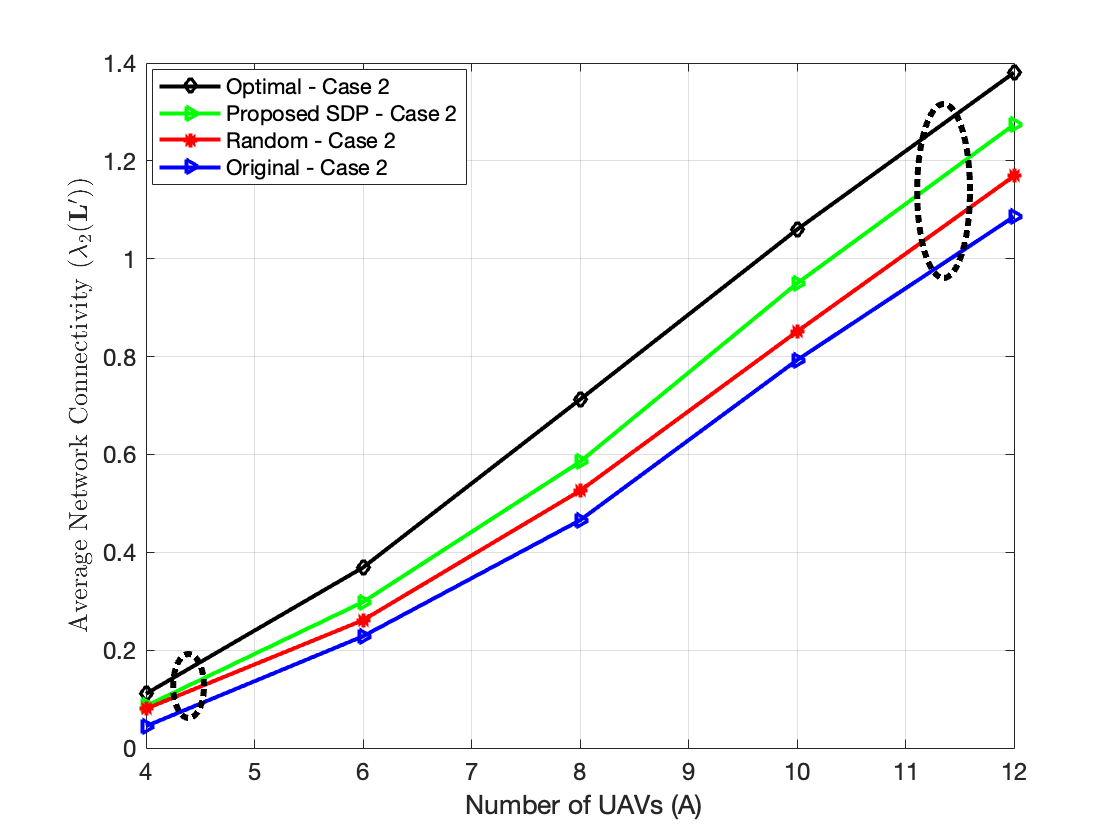}
				\caption{The average network connectivity $\lambda_2(\mathbf L')$ of case 2 versus the number of UAVs $A$.}
     \vspace{-5.5mm}
			\end{center}
   \label{fig2_case2}
		\end{figure}

In Figs. 2 and 3, we show the average network connectivity versus the number of UAVs $A$ for both cases. For a small number of UAVs in Figs. 2  and 3, the proposed optimal and SDP schemes offer a slight performance gain in terms of  network connectivity compared to the original and the random schemes. This is because our proposed schemes have a few options of links, where the RIS can direct the signal of the UE to a few number of UAVs. However, when the number of UAVs increases, the proposed schemes smartly selects an effective UE-RIS-UAV link  that significantly maximizes the network connectivity. It is noted that $\lambda_2(\mathbf L')$ of all schemes increases  with the number of UAVs since adding more connected nodes to the network increases the number of edges, which increases the network connectivity. It is also noted that the values of $\lambda_2(\mathbf L')$ in Fig. 3 are smaller than the values of  $\lambda_2(\mathbf L')$ in Fig. 2 for all the UAVs configurations. This is reasonably because the number of unconnected nodes that represent the UEs in Fig. 3 of case $2$, i.e., $U=10$,  is larger than those of  Fig. 2 of case $1$, which is one UE.  This makes the network of case $2$ less connected (i.e., more UE nodes and no links between them), thus low  network connectivity in Fig. 3. When $A>8$ in Fig. 3, the  average network connectivity of all the schemes increases   significantly with $A$, which follows the same behaviour of  Fig. 2 that is $A>U$.

\begin{figure}[t!]
			\begin{center}
				\includegraphics[width=0.95\linewidth, draft=false]{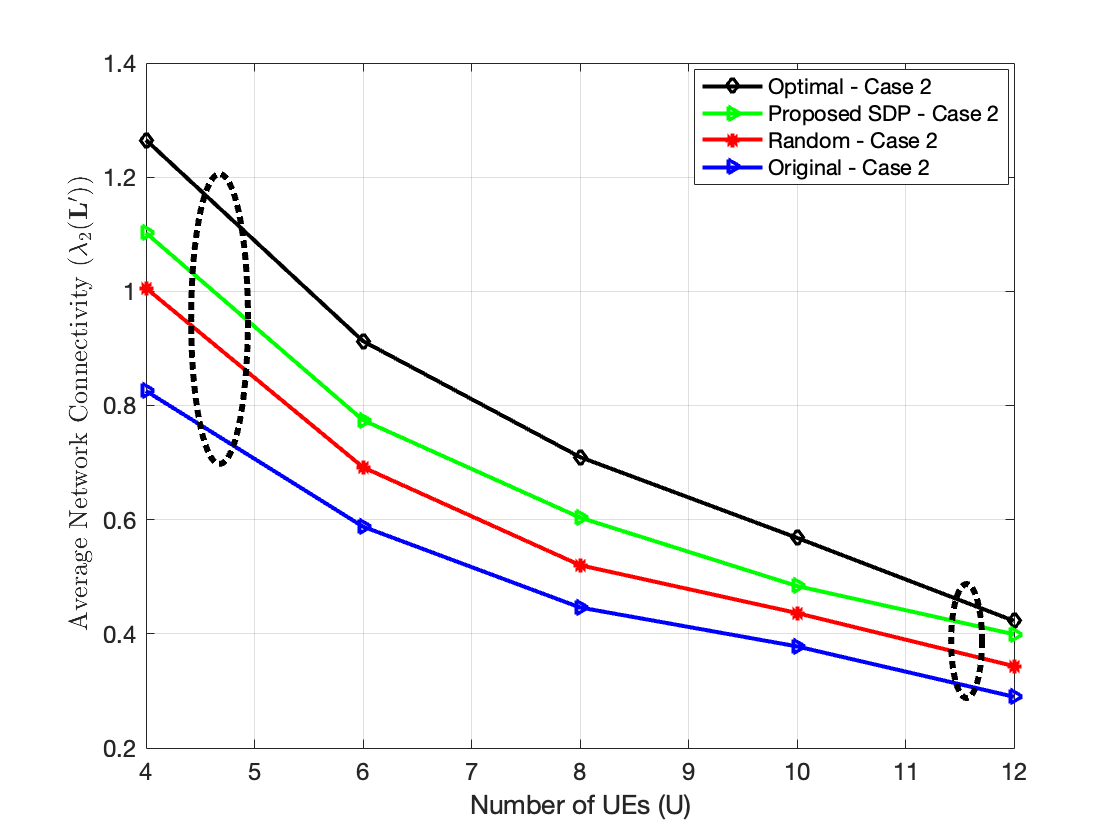}
				\caption{The average network connectivity $\lambda_2(\mathbf L')$ versus the number of UEs $U$.}
     \vspace{-5.5mm}
			\end{center}
   \label{fig3_case2}
		\end{figure}

\begin{figure}[t!]
			\begin{center}
				\includegraphics[width=0.95\linewidth, draft=false]{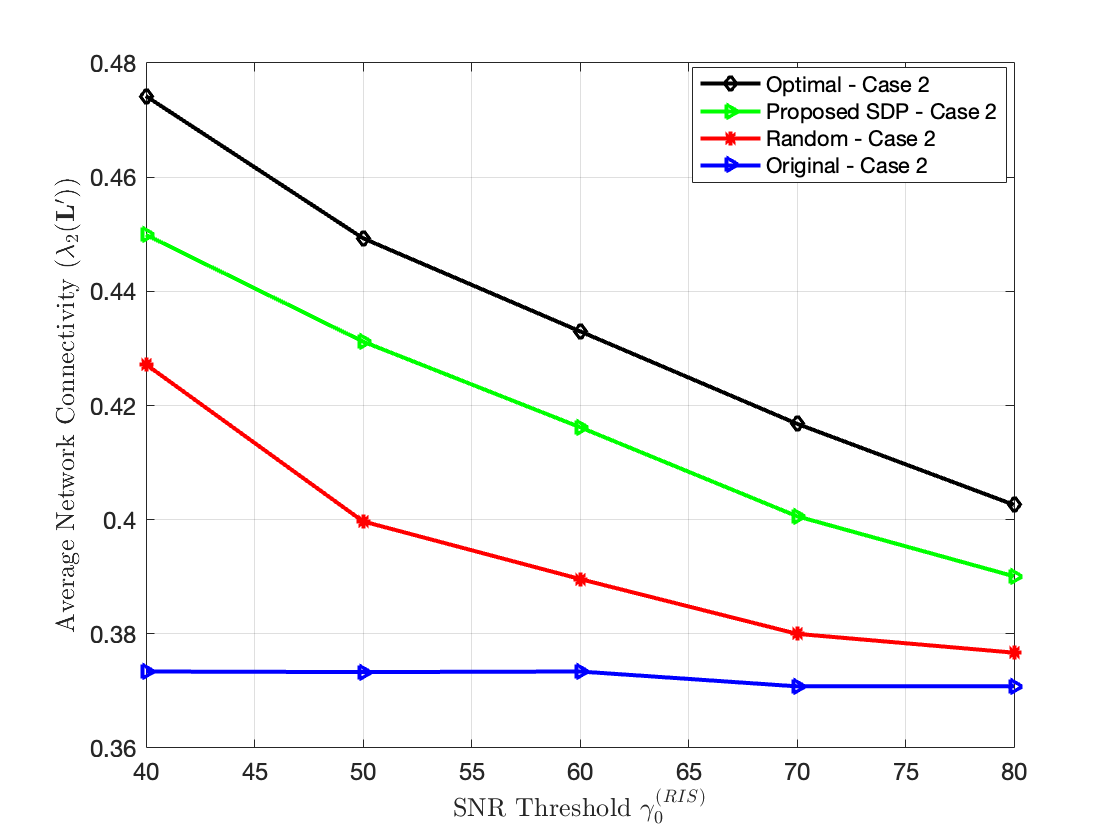}
				\caption{The average network connectivity $\lambda_2(\mathbf L')$ versus SNR threshold $\gamma_0^\text{(RIS)}$ in dB.}
     \vspace{-5.5mm}
			\end{center}
   \label{fig3_case2}
		\end{figure}

In Fig. 4, we plot the network connectivity versus the number of UEs $U$ for case $2$. From Fig. 4, we can see that the proposed SDP outperforms the original and the random schemes in terms of network connectivity. Notably, the network connectivity of all the schemes decreases as  the number of UEs increases, since adding more unconnected UEs may result in a sparse graph with low network connectivity.

In Fig. 5, we show the impact of the SNR threshold $\gamma_0^\text{(RIS)}$ on the network connectivity for case $2$. For small SNR threshold,   all the links between the UEs and the UAVs through the RIS can satisfy this SNR threshold, thus many alternative links between the potential UE and the UAVs to select to maximize the network connectivity. On the other hand, for high RIS SNR threshold, a few UE-RIS-UAV links can satisfy such high SNR threshold,  thus the network connectivity of all the schemes is degraded, and it becomes close to the original scheme, which does not get affected by changing $\gamma_0^\text{(RIS)}$.

It is worth remarking that while the random scheme adds a random link to the network, the original scheme does not add a link. The proposed solutions balance between the aforementioned aspects by judiciously selecting an effective link, between a UE and a UAV,  that maximizes the network connectivity. This utilizes the benefits of the cooperation between an appropriate scheduling algorithm design and RIS phase shift configurations.  Compared to the optimal scheme, our proposed SDP has a certain degradation in network connectivity that comes as the
achieved polynomial computational complexity as compared to the high complexity of the optimal scheme that is in the order of $\mathcal O(4 E' V^3/3)$ \cite{L}.

\section{Conclusion}\label{C}

In this paper, we proposed a novel joint UE-UAV scheduling and RIS phase shift optimization for achieving connected and resilient RIS-assisted UAV networks. We leveraged the RIS to add more links to the network by opportunistically reflecting the signal of the UE to the appropriate UAV such that the network connectivity is maximized. The problem of maximizing the network connectivity was 
formulated in two cases of a single UE and multiple UEs, and optimal and efficient SDP solutions were proposed for the two problem cases, respectively. Simulation results showed that both the proposed schemes result in improved  network connectivity as compared to the existing solutions. Such promising performance gain can be significantly improved for the case of multiple RISs, which will be pursued in our future work.

\end{document}